\documentclass[12pt]{article}
\pdfoutput=1
\usepackage{epsfig,eucal,bm,mathtools}
\usepackage[T1]{fontenc}
\usepackage[breaklinks=true,colorlinks=true,backref,pagebackref]{hyperref}

\usepackage{graphicx,epsfig}

 \usepackage{amsfonts,amssymb}
\usepackage{amsmath}
\usepackage{youngtab}

\usepackage{amsthm}
   \def\d{\delta}

\begin{document}

\title{Irreducible matrix resolution of the elasticity tensor for symmetry systems}
\author{Yakov Itin\\
  Inst. Mathematics, Hebrew Univ. of Jerusalem, \\
  Givat Ram,
  Jerusalem 91904, Israel,\\
  and Jerusalem College of Technology,
  Jerusalem 91160, Israel,\\
  email: itin@math.huji.ac.il
}

\date{\today, {\it file ElastMatrix1.tex}}
\maketitle

\begin{abstract} In linear elasticity, a fourth order elasticity (stiffness) tensor of 21 independent components completely describes deformation properties of a material. Due to Voigt, this tensor is conventionally represented by a  $6\times 6$ symmetric matrix. This classical matrix representation does not conform with the irreducible decomposition of the elasticity tensor. In this paper, we construct two alternative matrix representations. The $3\times 7$  matrix representation is   in a correspondence  with the permutation transformations of indices and with the general linear transformation of the basis. An additional representation of the  elasticity tensor by three $3\times 3$ matrices is suitable for description the irreducible decomposition under  the rotation transformations. We present the elasticity tensor of all crystal systems in these compact matrix forms and construct the hierarchy  diagrams based on this representation.

\end{abstract}

{\it {Key index words:}} anisotropic elasticity tensor,
  irreducible decomposition, matrix representation. 


\pagestyle{myheadings}\markright{Y. Itin\hfill Irreducible matrix resolution of the elasticity tensor\hfill}

\tableofcontents

\section{Introduction }

In anisotropic linear elasticity, the {\it  generalized Hooke law} \cite{Landau}, \cite{Love}, \cite{Nye}, postulates a linear
relation between  two symmetric 2nd order tensors, the stress tensor 
$\sigma^{ij}$ and the strain tensor $\varepsilon_{kl}$, 
\begin{equation}\label{hooke}
\sigma^{ij}=C^{ijkl}\,\varepsilon_{kl}
\end{equation}
with the 4th order {\it  elasticity (stiffness)  tensor} $C^{ijkl}$.
In 3-dimensional space, a generic 4th order tensor has a big set of 81 independent components. Since the elasticity tensor  is assumed to satisfy  the standard (minor and major) symmetry relations 
\begin{equation}\label{intro2}
C^{ijkl}=C^{jikl}=C^{ijlk},\qquad C^{ijkl}=C^{klij}\,,
\end{equation}
 it is left with 21 independent components only. 
Elasticity tensor  completely characterizes the deformed material in the linear regime. The components of $C^{ijkl}$ are refereed to as the  elasticity modules, the stiffness modules, or, merely,  as the  elasticities. With  the relativity small number of the independent components at hand, it is naturally to look for a  compact  matrix representation  of the elasticity tensor. The classical Voigt's $(6\times 6)$ symmetric matrix representation with 21 independent components is widely used.

But the specific entries of $C^{ijkl}$ are not really the intrinsic characteristics of the material  inasmuch as  they depend on the choice of the coordinate system. 
In order to deal with the proper material parameters, one must look for  {\it invariants of the elasticity  tensor,} see \cite{Norris}. As it is shown recently \cite{Quad}, \cite{Olive}, these invariant are connected to the unique irreducible decomposition of the elasticity tensor space into a direct sum of five invariant subspaces. When this   decomposition is applied to the  Voigt matrix, the corresponding five $(6\times 6)$ matrices do not show any systematic order.

In this paper, we are looking for an alternative matrix representation  of the elasticity tensor that is confirmed  with the irreducible decomposition. 
In Sect. 2, we present a representation with  $(3\times 7)$ matrix, which rows have similar permutation transformation behavior. This representation is useful for description the  decomposition of the elasticity tensor into Cauchy and non-Cauchy irreducible parts  \cite{Cauchy2}. 
But there is still a third  useful version of the matrix representation of elasticity tensor that we present in Sect. 3.  It includes two symmetric $(3\times 3)$-matrices of $6+6$ independent components and a generic matrix of 9 independent components. The symmetric matrices represent two tensors with proper transformation law. In particular, the   traces of these matrices are invariant. Our third generic matrix is not a tensor so it can be used  only as a compact representation of the corresponding irreducible part. 
In Sect. 4, we present the explicit expressions of the irreducible matrices for all symmetry classes of crystals. This {\it irreducible resolution} of  crystal's elasticity tensor   is used in  Sect. 5 for describing the hierarchy of the symmetry systems. This problem is usually studied   in literature by comparison the  groups of symmetry. In our construction, it is based on the inclusions of the irreducible parts. In most aspects, these two alternative approaches give    the equivalent results. Some small but important differences are   outlined. We also provide the Venn diagrams that  express the logical relationship between the symmetry classes. 
In Conclusion Sect., we discuss the problems of physical interpretation of the irreducible parts of the elasticity tensor. 

Notations: 
We use the standard tensor conventions that  distinguish between covariant and contravariant indices. The 3 dimensional indices are denoted by lower-case Latin letters, $i,j,\cdots=1,2,3.$ Upper-case Latin letters are used for 6 dimensional indices, $I,J,\cdots=1,\cdots, 6$. In these notations,  two repeated indices can appear only in up-down positions and summation  for only two such   repeated indices is assumed. 
The indices of a tensor can be raised/lowered by the use of the metric tensor  $g^{ij}$ and $g_{ij}$. For instance, the lower components of the elasticity tensor are defined as $C_{ijkl}=g_{ii'}g_{jj'}g_{kk'}g_{ll'}C^{i'j'k'l'}$. 
Since in  elasticity literature a simplified notation  is frequently used, we provide in certain  cases  the both notations   and relate the corresponding quantities by the sign $\stackrel{*}=$. Notice that this shorthand  notation is applicable only in Euclidean space endowed with rectangular coordinates. 
\section{Matrix representations}
For an isotropic material, the elasticity tensor is completely expressed by the second-order metric tensor and two scalars. In the  anisotropic case,   a representation of $C^{ijkl}$ with  smaller order tensors   cannot be reached. Although  the non-tensorial  {\it  matrix representations} are meaningful only in a  specific coordinate system,  they are  still  useful for many purposes, in particular, for a  classification of  elastic materials. 
\subsection{Voigt's representation}
The standard ``shorthand'' notation of $C^{ijkl}$ is due to Voigt, see \cite{Love}:  One identifies a
symmetric pair $\{ij\}$ of three-dimensional  indices with a multi-index $I$ which changes in 
the range from $1$ to $6$. This identification can not be canonical,  it is   chosen conventionally as 
\begin{equation}\label{voigt1} 11\to 1\,, \,22\to
2\,,\, 33\to 3\,,\, 23\to 4\,,\, 31\to 5\,,\, 12\to 6\,.
\end{equation}
With this notation, the elasticity tensor is expressed as a symmetric $6\times 6$ matrix $C^{IJ}$.  Observe that Voigt's ``shorthand'' notations (\ref{voigt1}) are  applicable only because the minor
symmetries $C^{ijkl}=C^{jikl}=C^{ijlk}$ are valid. Moreover, due to the major symmetry $C^{ijkl}=C^{klij}$, the $6\times 6$ matrix is symmetric, $C^{IJ}=C^{JI}$. Explicitly,
Voigt's  representation of the elasticity tensor reads
\begin{equation}\label{voigt2}
C^{ijkl}=\begin{bmatrix}
C^{1111} & C^{1122} & C^{1133} & C^{1123} & C^{1131} & C^{1112} \\
* & C^{2222} & C^{2233} & C^{2223} & C^{2231} & C^{2212} \\
* & * & C^{3333} & C^{3323} & C^{3331} & C^{3312} \\
* & * & * & C^{2323} & C^{2331} & C^{2312} \\
* & * & * & * & C^{3131} & C^{3112} \\
* & * & * & * & * & C^{1212}
     \end{bmatrix} = \begin{bmatrix}
  C^{11} & C^{12} & C^{13} & C^{14} & C^{15} & C^{16} \\
* & C^{22} & C^{23} & C^{24} & C^{25} & C^{26} \\
* & * & C^{33} & C^{34} & C^{35} & C^{36} \\
* & * & * & C^{44} & C^{45} & C^{46} \\
* & * & * & * & C^{55} & C^{56} \\
* & * & * & * & * & C^{66}\, \end{bmatrix}
\end{equation}
Here,  the equal components of the symmetric matrices are denoted by the star.  For a most general anisotropic material, such as a triclinic crystal, all the components explicitly displayed in (\ref{voigt2}) are nonzero and independent of one another.
An advantage of Voigt's matrix representation  is that it allows to write down the generalized Hook's law (\ref{hooke}) in a compact six-dimensional matrix form $\sigma^I=C^{IJ}\varepsilon_J$. A disadvantage of (\ref{voigt2}) that  this matrix representation mixes the components with similar permutation  properties. 
For instance, the components $C^{1122}$ and $C^{1212}$ that are related by a simple permutation of two indices are represented very differently: as $C^{12}$ and $C^{66}$,  respectively. 
As a result, the components that belong to certain irreducible part of the elasticity tensor are distributed almost randomly in the body of the  six-dimensional matrix. Also,  Nye's diagrams \cite{Nye}  that are designed to provide a systematic view of symmetries of elasticity tensor for specific crystal   systems are presented in a rather complicated  form in this standard $6\times 6$-matrix description. 
\subsection{An alternative representation}
In order to make   the permutation properties  of the elasticity tensor components visible, we construct an alternative matrix representation. 
We assemble 21 independent components of the tensor $C^{ijkl}$ into a  $3\times 7$ matrix:
\begin{equation}\label{new1}
C^{ijkl}=
\begin{bmatrix}
C^{1111}&C^{2222}&C^{3333}\\ 
C^{1112}&C^{2221}&C^{3331}\\ 
C^{1113}&C^{2223}&C^{3332}\\ 
C^{1122}&C^{1133}&C^{2233}\\ 
C^{1212}&C^{1313}&C^{2323}\\ 
C^{1123}&C^{2213}&C^{3312}\\ 
C^{1213}&C^{1223}&C^{1332}
\end{bmatrix}
\begin{bmatrix}
1\\2\\2\\2\\4\\4\\8
\end{bmatrix}
\equiv
\begin{bmatrix}
C^{11}&C^{22}&C^{33}\\ 
C^{16}&C^{26}&C^{35}\\ 
C^{15}&C^{24}&C^{34}\\ 
C^{12}&C^{13}&C^{23}\\
C^{66}&C^{55}&C^{44}\\ 
C^{14}&C^{25}&C^{36}\\ 
C^{56}&C^{46}&C^{45}
\end{bmatrix}.
\end{equation}
The rows of this matrix are  composed from the components with the similar permutation properties: In the first row, we choose the terms with four  identical indices; in the second and the third rows -- the terms with three identical indices. Then we give two rows of the terms each of which has two pairs of identical indices. In the last two rows, the terms with only one pair of identical indices are given. The second matrix of (\ref{new1}) represents the same terms in Voigt's  notation.

Notice the weight factors that must be adjoined to the  seven  rows of the matrix (\ref{new1}). 
The meaning of these factors is as follows: The entries of the first row are unique. 
Every entry  of the second, third and fourth rows  represents, in fact, two equal components, for instance,   $C^{1122}=C^{2211}$. The entries of the fifth and sixth rows represent four equal components, as $C^{1212}=C^{2112}=C^{1221}=C^{2121}$.  Every entry of the last row represents eight equal components, as 
 $C^{1213}=C^{1231}=C^{2113}=C^{2131}=C^{1312}=C^{1321}=C^{3112}=C^{3121}$. 
 The weight factors must be taken into account when one provides summation in the expressions including the elasticity tensor.
\section{Irreducible decomposition}
In this section, we provide matrix representations of the irreducible pieces of the elasticity tensor. The irreducible decomposition is constructed in two steps: We start with Young's decomposition relative to the group $S_4$ of permutations of four indices in $C^{ijkl}$. This decomposition is  equivalent to the irreducible decomposition relative to the group $GL(3,{\mathbb R})$ of general linear transformation of the basis. As a result, we obtain two $GL(3,{\mathbb R})$-irreducible parts of the elasticity tensor.  In the second step,  we decompose  these two parts successively  by extracting all possible  traces. This procedure requires metric tensor thus it yields the irreducible decomposition under the action of the rotation group $SO(3,{\mathbb R})$. Finally we obtain irreducible decomposition of the elasticity tensor into five independent pieces. For more details and formal proofs of the facts given here, see \cite{Cauchy2}. 
\subsection{$GL(3)$-decomposition}
 The irreducible decomposition of the elasticity tensor under the permutation group $S_4$  is described by two Young diagrams: 
\begin{equation}\label{irr3}
  \Yvcentermath1\yng(1) \otimes \yng(1)\otimes\yng(1)\otimes\yng(1)
  =\yng(4)\oplus \yng(2,2)\,
\end{equation}
Here the left-hand side represents the generic fourth order tensor as a tensor product of four basis vectors. On the right-hand side, the first diagram describes a totally symmetric tensor, while the second diagram describes a tensor that is partially symmetric and partially antisymmetric. The corresponding sub-tensors can be calculated by applying products of the symmetrization and antisymmetrization operators from group algebra of the group $S_4$ taken in some specific order. 
A straightforward calculation of such type terms can be rather complicated.   In our case with only two diagrams at hand, the first term is  a totally symmetric combination while the second term can be obtained merely as a residue. 
Consequently, the decomposition of the elasticity tensor reads 
\begin{equation}\label{decomp}
C^{ijkl}=S^{ijkl}+A^{ijkl}, 
\end{equation}
where the totally symmetric part reads
\begin{equation}\label{firsty}
S^{ijkl}=C^{(ijkl)}=\frac 13\left(C^{ijkl}+C^{iklj}+ C^{iljk}\right)
\end{equation}
and the residue $A^{ijkl}=C^{ijkl}-S^{ijkl}$ is given by
\begin{equation}\label{firsty'}
A^{ijkl}=\frac 13\left(2C^{ijkl}-C^{ilkj}-C^{iklj}\right). 
\end{equation}
This decomposition is irreducible and unique under the group $S_4$. It means that every action of additional symmetrization operators  on the tensors $S^{ijkl}$ and $A^{ijkl}$ preserves them or gives zero. Moreover, these  terms preserve their symmetry properties under arbitrary nondegenerate linear transformations of basis. 

As it was shown in \cite{Cauchy1}, the equation 
\begin{equation}\label{Cauchy-rel}
A^{ijkl}=0\end{equation}
is the irreducible invariant representation of the well-known {\it Cauchy relation.} Thus, we call $S^{ijkl}$ the {\it Cauchy part} and $A^{ijkl}$ the {\it non-Cauchy part} of the elasticity tensor. 

The partial tensors in Eq.(\ref{decomp}) satisfy the { minor and the major symmetries}
 \begin{equation}\label{alg1}
S^{ijkl}=S^{jikl}=S^{klij}\qquad {\rm and}\qquad  A^{ijkl}=A^{jik]}=A^{klij}\,.
\end{equation}
Thus, these two parts can serve as elasticities by themselves.

If we denote the vector space of the elasticity tensor by ${\cal C}$, the
irreducible decomposition (\ref{decomp}) signifies the reduction of ${\cal C}$
into the {\it direct sum of its two subspaces,} ${\cal S}\subset{\cal C}$ for the tensor $S$, and ${\cal A}\subset{\cal C}$ for the tensor $A$,
\begin{equation}\label{alg1}
{\cal C}={\cal S}\oplus{\cal A},\quad {\rm in\,\, particular,} \quad 
{\rm dim} \, {\cal C}=21\,,\qquad {\rm dim} \, {\cal S}=15\,,\qquad {\rm dim} \, {\cal A}=6\,.
\end{equation}

Moreover, the irreducible pieces $S^{ijkl}$ and $A^{ijkl}$ are orthogonal to one  another in the following sense: 
\begin{equation}\label{ORT1}
S^{ijkl}A_{ijkl}=0\,.
\end{equation}
Here the indices of $A^{ijkl}$ are lowered with the metric tensor. Consequently the standard 
 Pythagorean theorem holds:   The Euclidean (Frobenius)  squares of the tensors 
\begin{equation}\label{ORT2}
{\widetilde C}^2=C^{ijkl}C_{ijkl}\,,\qquad {\widetilde S}^2=S^{ijkl}S_{ijkl}\,,\qquad {\widetilde A}^2=A^{ijkl}A_{ijkl}
\end{equation}
 satisfy the relation 
\begin{eqnarray}\label{ORT3}
{\widetilde C}^2={\widetilde S}^2+{\widetilde A}^2 \,.
\end{eqnarray}

In Voigt's notations, the decomposition is presented as 
\begin{eqnarray}\label{voigt-dec}
&&C^{ijkl}=\begin{bmatrix}
C^{1111} & C^{1122} & C^{1133} & C^{1123} & C^{1131} & C^{1112} \\
* & C^{2222} & C^{2233} & C^{2223} & C^{2231} & C^{2212} \\
* & * & C^{3333} & C^{3323} & C^{3331} & C^{3312} \\
* & * & * & C^{2323} & C^{2331} & C^{2312} \\
* & * & * & * & C^{3131} & C^{3112} \\
* & * & * & * & * & C^{1212}
     \end{bmatrix}=\nonumber\\
     &&
     \begin{bmatrix}
\pmb{S^{1111}} & \pmb{S^{1122}} & \pmb{S^{1133}} & \pmb{S^{1123}} & \pmb{S^{1131}} & \pmb{S^{1112}} \\
* & \pmb{S^{2222}} & \pmb{S^{2233}} & \pmb{S^{2223}} & \pmb{S^{2231}} & \pmb{C^{2212}} \\
* & * & \pmb{S^{3333}} & \pmb{S^{3323}} & \pmb{S^{3331}} & \pmb{S^{3312}} \\
* & * & * & S^{2323} & S^{2331} & S^{2312} \\
* & * & * & * & S^{3131} & S^{3112} \\
* & * & * & * & * & S^{1212}
     \end{bmatrix}+
     \begin{bmatrix}
0 & \pmb {A^{1122}} & \pmb {A^{1133}} & \pmb {A^{1123}} & 0 & 0 \\
* & 0 &\pmb { A^{2233}} & 0 & \pmb {A^{2231}} & 0 \\
* & * & 0 & 0 & 0 & \pmb {A^{3312}} \\
* & * & * & A^{2323} & A^{2331} & A^{2312} \\
* & * & * & * & A^{3131} & A^{3112} \\
* & * & * & * & * & A^{1212}
     \end{bmatrix}\nonumber
     \end{eqnarray}
 We use here the bold font to denote the independent components into the three first rows of the matrices.  In this representation, the location of equal components and  zeros   do not show any order.     

We present now the decomposition (\ref{decomp}) in  term of $3\times7$ matrices. Calculating with (\ref{firsty}) we derive the symmetric Cauchy part
\begin{equation}\label{Cauchy1}
S^{ijkl}=\begin{bmatrix}
S^{1111}&S^{2222}&S^{3333}\\ 
S^{1112}&S^{2221}&S^{3331}\\ 
S^{1113}&S^{2223}&S^{3332}\\ 
S^{1122}&S^{1133}&S^{2233}\\ 
S^{1212}&S^{1313}&S^{2323}\\ 
S^{1123}&S^{2213}&S^{3312}\\ 
S^{1213}&S^{1223}&S^{1332}
\end{bmatrix}
=
\frac13\begin{bmatrix}
3\pmb {C^{11}}&3\pmb {C^{22}}&3\pmb {C^{33}}\\ 
3\pmb {C^{16}}&3\pmb {C^{26}}&3\pmb {C^{35}}\\ 
3\pmb {C^{15}}&3\pmb {C^{24}}&3\pmb {C^{34}}\\ 
\pmb {C^{12}+2C^{66}}&\pmb{C^{13}+2 C^{55}}&\pmb {C^{23}+2C^{44}}\\
C^{12}+2C^{66}&C^{13}+2C^{55}&C^{23}+2C^{44}\\
\pmb {C^{14}+2C^{56}}&\pmb {C^{25}+2C^{46}}&\pmb {C^{36}+2C^{45}}\\
C^{14}+2C^{56}&C^{25}+2C^{46}&C^{36}+2C^{45}
\end{bmatrix}.
\end{equation}
 Here, 15  independent  components of the  Cauchy part $S^{ijkl}$ are  visibly expressed in the bold font. The additional  6 dependent components are located close to their equal entries. 

Similarly, the  non-Cauchy part (\ref{firsty'}) is given by
\begin{equation}\label{nCauchy1}
A^{ijkl}=\begin{bmatrix}
A^{1111}&A^{2222}&A^{3333}\\ 
A^{1112}&A^{2221}&A^{3331}\\ 
A^{1113}&A^{2223}&A^{3332}\\ 
A^{1122}&A^{1133}&A^{2233}\\ 
A^{1212}&A^{1313}&A^{2323}\\ 
A^{1123}&A^{2213}&A^{3312}\\ 
A^{1213}&A^{1223}&A^{1332}
\end{bmatrix}
=
\frac13\begin{bmatrix}
0&0&0\\ 
0&0&0\\ 
0&0&0\\ 
2\pmb {\left(C^{12}-C^{66}\right)}&2\pmb {\left(C^{13}-C^{55}\right)}&2\pmb {\left(C^{23}-C^{44}\right)}\\
-\left(C^{12}-C^{66}\right)&-\left(C^{13}-C^{55}\right)&-\left(C^{23}-C^{44}\right)\\ 
2\pmb {\left(C^{14}-C^{56}\right)}&2\pmb {\left(C^{25}-C^{46}\right)}&2\pmb {\left(C^{36}-C^{45}\right)}\\
-\left(C^{14}-C^{56}\right)&-\left(C^{25}-C^{46}\right)&-\left(C^{36}-C^{45}\right)
\end{bmatrix}.
\end{equation}
We  recognize here 6 independent components and 6 their dependent copartners.  Moreover, the identity 
\begin{equation}
    A^{ijkl}+A^{iljk}+A^{iklj}=0\,
\end{equation}
can be checked immediately with the explicit expression in the right hand side of (\ref{nCauchy1}). 

With  account of the factors given in (\ref{new1}), we can straightforwardly prove also the orthogonality relation (\ref{ORT1})  from this representation. It is enough to check the  term-by-term products of the matrices (\ref{Cauchy1}) and (\ref{nCauchy1}). 
\subsection{$SO(3,{\mathbb R})$-decomposition}
The group of rotations $SO(3,{\mathbb R})$ is a subgroup of the general linear group $GL(3)$ that preserves the scalar product. Under the action of this subgroup, two  $GL(3)$-irreducible pieces of the  elasticity tensor are decomposed successively in smaller irreducible parts. Since  $SO(3,{\mathbb R})$ group brings in consideration  only one new object, the metric tensor $g_{ij}$, the new irreducible parts are determined by  contractions of the elasticity tensor with  $g_{ij}$. In   matrix description, these contractions are treated as traces. 
\subsubsection{Cauchy part}
Due to the total symmetry of  $S^{ijkl}$, all possible contraction of it with  $g^{ij}$ are equal to one another, thus we have only one  symmetric second-order tensor 
 \begin{equation}\label{S2}
   S^{ij}:=S^{ijkl}g_{kl}
   =\frac 13 \left(C^{ijkk}+2C^{ikkj}\right)
   \end{equation}
   and one scalar 
 \begin{equation}\label{S2a}
   S:=S^{ij}g_{ij}
   =\frac 13 \left(C^{iikk}+2C^{ikki}\right)\,.
\end{equation}
We define by $P^{ij}$ the traceless part of the tensor $S^{ij}$ 
\begin{equation}\label{S00} P^{ij}:=S^{ij}-\frac 13
  Sg^{ij}\,,\qquad\text{with}\qquad g_{ij}P^{ij}=0\,.
\end{equation}
With these notations, we are able to define the first (scalar) part of $S^{ijkl}$ as
 \begin{equation}\label{sub2}
  {}^{(1)}\!S^{ijkl}=\frac S5 g^{(ij}g^{kl)}=\frac S{15}\left(g^{ij}g^{kl} +g^{ik}g^{jl}+g^{il}g^{jk}\right).
  \end{equation}
 The second part of $S^{ijkl}$ is defined as 
  \begin{equation}\label{sub2aa}
  {}^{(2)}\!S^{ijkl}=
\frac 67 P^{(ij}g^{kl)}=\frac 17\left(P^{ij}g^{kl}+P^{ik}g^{jl}+P^{il}g^{jk}+P^{jk}g^{il}+P^{jl}g^{ik}+P^{kl}g^{ij}\right),
\end{equation}
and satisfies the relation ${}^{(2)}\!S^{ijkl}g_{ij}g_{kl}=0$. 
The choice of the leading coefficients in (\ref{sub2}, \ref{sub2aa}) guarantees the  residue  part,   
\begin{equation}\label{sub1x}
   {}^{(3)}\!S^{ijkl}=S^{ijkl}-{}^{(1)}\!S^{ijkl}-{}^{(2)}\!S^{ijkl}=R^{ijkl},
\end{equation}
to be  traceless and totally symmetric:  
\begin{equation}\label{sub1y}
   {}^{(3)}\!S^{ijkl}={}^{(3)}\!S^{(ijkl)}\,,\qquad {}^{(3)}\!S^{ijkl}g_{ik}=0\,.
\end{equation}

In $(3\times 7)$-matrix notation, the first scalar piece of the Cauchy part reads
\begin{equation}\label{S1}
{}^{(1)}S^{ijkl}=\begin{bmatrix}
{}^{(1)}S^{1111}&{}^{(1)}S^{2222}&{}^{(1)}S^{3333}\\ 
{}^{(1)}S^{1112}&{}^{(1)}S^{2221}&{}^{(1)}S^{3331}\\ 
{}^{(1)}S^{1113}&{}^{(1)}S^{2223}&{}^{(1)}S^{3332}\\ 
{}^{(1)}S^{1122}&{}^{(1)}S^{1133}&{}^{(1)}S^{2233}\\ 
{}^{(1)}S^{1212}&{}^{(1)}S^{1313}&{}^{(1)}S^{2323}\\ 
{}^{(1)}S^{1123}&{}^{(1)}S^{2213}&{}^{(1)}S^{3312}\\ 
{}^{(1)}S^{1213}&{}^{(1)}S^{1223}&{}^{(1)}S^{1332}
\end{bmatrix}
=
\frac 1{15} S\begin{bmatrix}
3&3&3\\ 
0&0&0\\ 
0&0&0\\ 
1&1&1\\
1&1&1\\ 
0&0&0\\ 
0&0&0\\ 
\end{bmatrix}.
\end{equation}
The second part with 5 independent components is given by 
\begin{equation}\label{S2aa}
{}^{(2)}S^{ijkl}=\begin{bmatrix}
{}^{(2)}S^{1111}&{}^{(2)}S^{2222}&{}^{(2)}S^{3333}\\ 
{}^{(2)}S^{1112}&{}^{(2)}S^{2221}&{}^{(2)}S^{3331}\\ 
{}^{(2)}S^{1113}&{}^{(2)}S^{2223}&{}^{(2)}S^{3332}\\ 
{}^{(2)}S^{1122}&{}^{(2)}S^{1133}&{}^{(2)}S^{2233}\\ 
{}^{(2)}S^{1212}&{}^{(2)}S^{1313}&{}^{(2)}S^{2323}\\ 
{}^{(2)}S^{1123}&{}^{(2)}S^{2213}&{}^{(2)}S^{3312}\\ 
{}^{(2)}S^{1213}&{}^{(2)}S^{1223}&{}^{(2)}S^{1332}
\end{bmatrix}
=
\frac17\begin{bmatrix*}[r]
6\pmb {P^{11}}&6\pmb {P^{22}}&6P^{33}\\ 
3\pmb {P^{12}}&3P^{12}&3\pmb {P^{13}}\\ 
3P^{13}&3\pmb {P^{23}}&3P^{23}\\ 
-P^{33}&-P^{22}&-P^{11}\\
-P^{33}&-P^{22}&-P^{11}\\
\,\,\,\,P^{23}&P^{13}&P^{12}\\
P^{23}&P^{13}&P^{12}\\
\end{bmatrix*}.
\end{equation}
Here we emphasized 5 independent components. 
Recall that $P^{ij}$ is a traceless matrix, $P^{11}+{P^{22}}+P^{33}=0$. 
The orthogonal property 
\begin{equation}\label{ort}
{}^{(1)}S_{ijkl}{}^{(2)}S^{ijkl}=0
\end{equation}
follows straightforwardly from the matrix representations (\ref{S1}, \ref{S2aa}). 
Consequently, the second piece of the Cauchy part is completely described by a symmetric traceless matrix of 5 independent components 
\begin{equation}\label{S2}
P^{ij}=\begin{bmatrix}
P^{11}&P^{12}&P^{13}\\
*&P^{22}&P^{23}\\
*&*&P^{33}
\end{bmatrix},\qquad {\rm where} \qquad 
P^{11}+P^{22}+P^{33}=0.
\end{equation}
The third  part  ${}^{(3)}S^{ijkl}$ has nine independent components. Thus it can be represented by an  asymmetric $3\times 3$ matrix which we denote by $R^{ij}$.
We introduce a following parametrization: 
\begin{equation}\label{R-def}
{}^{(3)}S^{iiij}\stackrel{*}=R^{ij}\qquad\qquad \mbox{no summation!}
\end{equation}
For example, $R^{1111}=R^{11}$, $R^{1112}=R^{12}$, $R^{2221}=R^{21}$ etc. Using the traceless identity ${}^{(3)}S^{ijkl}g_{kl}=0$,  we express all components of the tensor ${}^{(3)}S^{ijkl}$ by linear combinations of the entries  of the matrix $R^{ij}$
\begin{equation}\label{S3}
{}^{(3)}S^{ijkl}\!=\!\begin{bmatrix}
{}^{(3)}S^{1111}&{}^{(3)}S^{2222}&{}^{(3)}S^{3333}\\ 
{}^{(3)}S^{1112}&{}^{(3)}S^{2221}&{}^{(3)}S^{3331}\\ 
{}^{(3)}S^{1113}&{}^{(3)}S^{2223}&{}^{(3)}S^{3332}\\ 
{}^{(3)}S^{1122}&{}^{(3)}S^{1133}&{}^{(3)}S^{2233}\\ 
{}^{(3)}S^{1212}&{}^{(3)}S^{1313}&{}^{(3)}S^{2323}\\ 
{}^{(3)}S^{1123}&{}^{(3)}S^{2213}&{}^{(3)}S^{3312}\\ 
{}^{(3)}S^{1213}&{}^{(3)}S^{1223}&{}^{(2)}S^{1332}
\end{bmatrix}
=
\frac 12\begin{bmatrix}
2\pmb {R^{11}}&2\pmb {R^{22}}&2\pmb {R^{33}}\\ 
2\pmb {R^{12}}&2\pmb {R^{21}}&2\pmb {R^{31}}\\ 
2\pmb {R^{13}}&2\pmb {R^{23}}&2\pmb {R^{32}}\\ 
R^{33}\!-\!R^{11}\!-\!R^{22}& R^{22}\!-\!R^{11}\!-\!R^{33}&R^{11}\!-\!R^{22}\!-\!R^{33}\\
R^{33}\!-\!R^{11}\!-\!R^{22}& R^{22}\!-\!R^{11}\!-\!R^{33}&R^{11}\!-\!R^{22}\!-\!R^{33}\\
-2\left(R^{23}+R^{32}\right)&-2\left(R^{13}+R^{31}\right)&-2\left(R^{12}+R^{21}\right)\\
-2\left(R^{23}+R^{32}\right)&-2\left(R^{13}+R^{31}\right)&-2\left(R^{12}+R^{21}\right)
\end{bmatrix}.
\end{equation}
Thus the third piece of the Cauchy part is completely characterized by the matrix 
\begin{equation}\label{R}
R^{ij}=\begin{bmatrix}
R^{11}&R^{12}&R^{13}\\
R^{21}&R^{22}&R^{23}\\
R^{31}&R^{32}&R^{33}
\end{bmatrix}.
\end{equation}
Notice a principle difference between the matrices $P^{ij}$ and $R^{ij}$. The former is defined by a covariant equation (\ref{S00}) thus it must be considered as a matrix representation of a tensor with a proper transformation law. Alternatively, $R^{ij}$ is  defined by a non-covariant equation (\ref{R-def}). So it is only a matrix but not a tensor. In particular, its trace does not have an invariant sense.  Hence the matrix $R^{ij}$ is meaningful only in a chosen coordinate system. Even with this restriction,  the representation $R^{ij}$ is useful for analysis of the elasticity tensor structure. In particular, since all components of the 9 dimensional tensor ${}^{(3)}S^{ijkl}$ are expressed linearly by 9 components of the matrix $R^{ij}$ these 9 components indeed form a basis of the tensor space. 
Moreover, if $R^{ij}=0$ in some coordinate system, then ${}^{(3)}S^{ijkl}=0$. Since it is a tensor equality we have  $R^{ij}=0$ in an  arbitrary coordinate system. 

With $(3\times 7)$-matrix representations and  the weight factors given in  (\ref{new1}), we can check straightforwardly  the orthogonality conditions
\begin{equation}\label{ort4}
{}^{(1)}S_{ijkl}{}^{(3)}S^{ijkl}=0\qquad {\rm and} \qquad {}^{(2)}S_{ijkl}{}^{(3)}S^{ijkl}=0\,.
\end{equation}
\subsubsection{Non-Cauchy part}
The  non-Cauchy piece $A^{ijkl}$  has 6 independent components. It is naturally to look for its  representation by a symmetric 2nd-order tensor. 
Using Levi-Civita's permutation tensor  $\epsilon_{ijk}=0,\pm 1$,  we define a   tensor
\begin{equation}\label{Del}
  \Delta_{mn}=\frac 13\epsilon_{mil} \epsilon_{njk}A^{ijkl}\,,
  \end{equation}
  which is symmetric $\Delta_{mn}=\Delta_{nm}$. 
  With the standard identity $\epsilon_{ijk}\epsilon^{imn}=(1/2)(\d^j_m\d^k_n-d^j_n\d^k_m)$, we can reverse  relation (\ref{Del}) and derive 
 \begin{equation}\label{Del1} A^{ijkl}=\frac 12\left(\epsilon^{imk}\epsilon^{ljn}-\epsilon^{iml}\epsilon^{kjn}\right)\Delta_{mn}\,.
\end{equation}
Thus the fourth-order tensor $A^{ijkl}$ and the second-order tensor $\Delta_{mn}$ are completely equivalent to one  another. 
We define a scalar $A$ that is a trace of the matrix $\Delta_{mn}$
\begin{equation}\label{sub6a}
 A=g^{mn}\Delta_{mn}\stackrel{*}=\frac 1{3}\left(C_{iikk}-C_{ikik}\right). 
\end{equation} 
Now the  tensor  $\Delta_{ij}$ can be  decomposed uniquely into the scalar and  traceless pieces:
\begin{equation}\label{sub6}
  \Delta_{ij}=Q_{ij}+\frac 13A  g_{ij}\,,\qquad {\rm with}\qquad Q_{ij}g^{ij}=0\,.
\end{equation}
Consequently, we have 
the non-Cauchy part decomposed  into two pieces 
\begin{equation}\label{sub7}
A^{ijkl}={}^{(1)}\!A^{ijkl}+{}^{(2)}\!A^{ijkl}\,.
\end{equation}
The scalar and the tensor  parts are given by
\begin{equation}\label{sub8a}
{}^{(1)}\!A^{ijkl}=\frac 16
A\left(2g^{ij}g^{kl}-g^{il}g^{jk}-g^{ik}g^{jl}\right)\,
\end{equation}
and
\begin{equation}\label{sub8b}
{}^{(2)}\!A^{ijkl}=\frac 12\left(\epsilon^{ikm}\epsilon^{jln}+
\epsilon^{ilm}\epsilon^{jkn}\right)Q_{mn} \,.
\end{equation}
The decomposition (\ref{sub7}) is unique, invariant, and irreducible under the action of the rotation group $SO(3,{\mathbb R})$. 

In terms of the $(3\times 7)$-matrix representation, we have
\begin{equation}\label{nCauchy2}
{}^{(1)}\!A^{ijkl}=\frac16 A\begin{bmatrix*}[r]
0&0&0\\ 
0&0&0\\ 
0&0&0\\ 
2&2&2\\
-1&-1&-1\\ 
0&0&0\\
0&0&0
\end{bmatrix*}\,,\qquad {}^{(2)}\!A^{ijkl}=
\frac12 \begin{bmatrix*}[r]
0&0&0\\ 
0&0&0\\ 
0&0&0\\ 
2 {Q_{33}}&2\pmb {Q_{22}}&2\pmb {Q_{11}}\\
-Q_{33}&-Q_{22}&-Q_{11}\\ 
-2Q_{23}&-2Q_{13}&-2Q_{12}\\
\pmb {Q_{23}}&\pmb {Q_{13}}&\pmb {Q_{12}}
\end{bmatrix*}.
\end{equation}
Here we emphasized five independent components of the matrix $Q_{ij}$. 
With the explicit form (\ref{nCauchy2}), we  straightforwardly  
 prove   the orthogonality relation
\begin{equation}\label{ort5}
 {}^{(1)}\!A_{ijkl}{}^{(2)}\!A^{ijkl}=0\,.
\end{equation}

Consequently, the second piece of the non-Cauchy part is completely described by a symmetric traceless matrix 
\begin{equation}\label{S2}
Q_{ij}=\begin{bmatrix}
Q_{11}&Q_{12}&Q_{13}\\
*&Q_{22}&Q_{23}\\
*&*&Q_{33}
\end{bmatrix}\,,\qquad Q_{11}+Q_{22}+Q_{33}=0.
\end{equation}

\subsection{Irreducible decomposition}
Let us collect our  results concerning  irreducible decomposition of the elasticity tensor into the sum of five parts
\begin{equation}\label{sub10b}
{C^{ijkl}=\sum^5_{A=1}{}^{(A)}C^{ijkl}=
 \left({}^{(1)}\!S^{ijkl}+{}^{(2)}\!S^{ijkl}+{}^{(3)}\!S^{ijkl}\right)+
\left({}^{(1)}\!A^{ijkl}+{}^{(2)}\!A^{ijkl}\right)}\,.
\end{equation}  
This decomposition is unique and irreducible under the action of the permutation group $S_4$. Uniqueness  means that any alternative procedure of construction the decomposition must give the same expressions for ${}^{(A)}C^{ijkl}$. The irreducibility yields that any additional symmetrization (or anti-symmetrization) of a specific piece in (\ref{sub10b}) preserves it or gives zero. 
When a rotation of the coordinate basis is applied, every piece ${}^{(A)}C^{ijkl}$ transforms into a terms with the same permutation symmetry. It means that the decomposition (\ref{sub10b}) is irreducible under the action of the rotation group  $SO(3,{\mathbb R})$. 

The decomposition (\ref{sub10b}) corresponds to the {\it direct sum decomposition} of the vector space of the elasticity tensor into five subspaces
\begin{equation}\label{ORT3}
{\cal C}= \left({}^{(1)}{\cal C}\oplus{}^{(2)}{\cal C}\oplus{}^{(3)}{\cal C}\right)\oplus \left({}^{(4)}{\cal C}\oplus{}^{(5)}{\cal C}\right),
\end{equation}
with the dimensions
\begin{equation}\label{ORT3a}
21= \left(1\,\,\oplus\, \,5\,\,\oplus\, \,9\right) \,\,\oplus\,\left(1\,\,\oplus \,\,5\right).
\end{equation}
It means that these subspaces do not have non-zero common tensors in their  intersections. Additionally, from the direct sum decomposition follows that an arbitrary tensor can be written as a unique  linear combination of the tensors laying in corresponding subspaces.   

When the Euclidean (Frobenius) scalar product of the tensors is involved, the irreducible pieces turn out to be  orthogonal to one  another: For $A, B=1,\cdots 5$ with $A\ne B$
\begin{equation}\label{ORT2}
{}^{(A)}C_{ijkl}{}^{(B)}C^{ijkl}=0\,.
\end{equation}
Consequently, the Euclidean squares, $C^2:=C_{ijkl}C^{ijkl}$ and ${}^{(A)}C^2:={}^{(A)}C_{ijkl}\,{}^{(A)}C^{ijkl}$ with $A=1,\cdots,5$, fulfill the ``Pythagorean theorem"
\begin{equation}\label{sub10a}
C^2=
\left({}^{(1)}C^2+{}^{(2)}C^2+{}^{(3)}C^2\right)+\left({}^{(4)}C^2+
{}^{(5)}C^2\right).
\end{equation}

The decomposition (\ref{sub10b}) is constructed from two scalars $S$ and $A$, two second-order traceless tensors $P_{ij}$ and $Q_{ij}$, and a  totally traceless and a  completely symmetric 4th order tensor $R_{ijkl}$. 
Tensors of the same types   emerge in the harmonic decomposition that is widely used in elasticity literature, see for instance \cite{Backus}, \cite{Baerheim}, \cite{Cowin1989}, \cite{Forte1}, \cite{Forte2} and  \cite{Forte3}. 
The harmonic  decomposition is generated from  the harmonic polynomials, i.e.,  the polynomial solutions of  Laplace's equation. The corresponding tensors are restricted to be completely symmetric and totally traceless. 
The most compact expression of this type was proposed by Cowin \cite{Cowin1989},  
\begin{eqnarray}\label{id1}
C^{ijkl}&=&ag^{ij}g^{kl}+b\left(g^{ik}g^{jl}+g^{il}g^{jk}\right)
+\left(g^{ij}{\hat A}^{kl}+g^{kl}{\hat A}^{ij}\right) +\nonumber\\
&&\left(g^{ik}{\hat B}^{jl}+g^{il}{\hat B}^{jk}+g^{jk}{\hat B}^{il}+g^{ik}{\hat B}^{jl}\right)+Z^{ijkl}\,.
\end{eqnarray}
It is straightforwardly to show that five  individual terms in this expression are reducible and do not represent elasticities by themselves. 

An alternative expression of Backus \cite{Backus}, \cite{Baerheim},
\begin{eqnarray}\label{id2}
C^{ijkl}&=&H^{ijkl}+ \left(H^{ij}g^{kl}+H^{ik}g^{jl}+H^{il}g^{jk}+H^{jk}g^{il}
+H^{jl}g^{ik}+H^{kl}g^{ij}\right)+\nonumber\\
&&H\left(g^{ij}g^{kl}+g^{ik}g^{jl}+g^{il}g^{jk}\right)+\nonumber\\
&&
\left(h^{ij}g^{kl}+h^{kl}g^{ij}-\frac 12 h^{jl}g^{ik}-
\frac 12h^{ik}g^{jl}-\frac 12h^{jk}g^{il} -\frac 12h^{il}g^{jk} \right)
+\nonumber\\
&&
h\left(g^{ij}g^{kl}-\frac12g^{il}g^{jk}-\frac12 g^{ik}g^{jl}\right)
\end{eqnarray}
is, in fact, equivalent to ours. 
Indeed, we immediately identify 
\begin{eqnarray}\label{id6}
{}^{(1)}\!S^{ijkl}&=&H\left(g^{ij}g^{kl}+g^{ik}g^{jl}+
g^{il}g^{jk}\right), \label{id9} \\
{}^{(2)}\!S^{ijkl}&=& H^{ij}g^{kl}+H^{ik}g^{jl}+H^{il}g^{jk}+H^{jk}g^{il}
+H^{jl}g^{ik}+H^{kl}g^{ij}\,, \label{id8} \\
{}^{(3)}\!S^{ijkl}&=& H^{ijkl}\,, \label{id7}\\ 
{}^{(1)}\!A^{ijkl}&=& h\left(g^{ij}g^{kl}-\frac12g^{il}g^{jk}-\frac12 g^{ik}g^{jl}\right), \\
{}^{(2)}\!A^{ijkl}&=&h^{ij}g^{kl}+h^{kl}g^{ij}-\frac 12 h^{jl}g^{ik}-
\frac 12h^{ik}g^{jl}-\frac 12h^{jk}g^{il} -\frac 12h^{il}g^{jk} \, \label{id10}
\end{eqnarray}
with the relations 
\begin{equation}\label{id11}
H=\frac 1{15} S\,,\qquad H^{ij}=\frac 17 P^{ij}\,, \qquad H^{ijkl}=R^{ijkl}\,,
\end{equation}
and
\begin{equation}\label{id12}
h=\frac 1{3} A\,,\qquad h^{ij}=-2 Q^{ij}\,.
\end{equation}
Consequently, our results concerning uniqueness,  irreducibility and direct sum decomposition valid also for  harmonic decomposition in Baskus's form. 
Moreover,  with the $S_4$ decomposition we are able to demonstrate the  origin of the difference between two scalars and two tensors  in (\ref{id6}).   The scalar $S$ and the tensor $P^{ij}$ come from the Cauchy part, so also $H$ and $H^{ij}$.  The scalar $A$ and the tensor $Q^{ij}$ come from the  non-Cauchy part, so also $h$ and $H^{ij}$. 
The difference between Cauchy and non-Cauchy scalars turns out  to be important in acoustic waves propagation \cite{Cauchy2}.

\section{Crystal systems}\label{Crystal}
We have shown that an arbitrary elasticity tensor can be expressed by two scalars $S$ and $A$, two tensors $P^{ij}$ and $Q^{ij}$, and a  matrix  $R^{ij}$. We study now how these objects can be used for characterization of the crystal materials belonging to different symmetry systems.   We use \cite{Landau}, \cite{Love}, and \cite{Haus} as our basic references for symmetry systems conventions.

\subsection{Triclinic system}
Triclinic system exhibits the most general anisotropy inasmuch as a triclinic crystal has no symmetry axes or mirror planes.  An only symmetry  of $C^{ijkl}$ in this system is  the central point transformation $x^i\to -x^i$ for all $i=1,2,3$. This symmetry holds  due to the even order of the elasticity tensor.    
The $(6\times 6)$-matrix representation  and $(3\times 7)$-matrix representation for this system are presented in (\ref{voigt2}) and (\ref{new1}), respectively. In this section, we present the expressions for the scalars $S$ and $A$, the symmetric matrices $P^{ij}$, $Q^{ij}$,  and for the asymmetric matrix $R^{ij}$ in terms of the components $C^{IJ}$.  

For the Cauchy part, we have   from (\ref{S2a}) the scalar in the form 
\begin{equation}\label{S1-scalar}
{}^{\rm (tr)}S=C^{11}+C^{22}+C^{33}+(2/3) \left(C^{12}+C^{13}+C^{23}+2C^{44}+2C^{55}+2C^{66}\right)\,.
\end{equation}
The symmetric traceless tensor (\ref{S00}) has  5 independent components, that are  expressed as
\begin{eqnarray}
{}^{\rm (tr)}P^{11}&=&C^{11}+(1/3) \left(C^{12}+C^{13}+2C^{55}+2C^{66}\right)-(1/3)S,\\
{}^{\rm (tr)}P^{22}&=&C^{22}+(1/3) \left(C^{12}+C^{23}+2C^{44}+2C^{66}\right)-(1/3)S,\\
{}^{\rm (tr)}P^{33}&=&C^{33} +(1/3) \left(C^{13}+C^{23}+2C^{44}+2C^{55}\right)-(1/3)S,  \\
{}^{\rm (tr)}P^{12}&=&C^{16}+C^{26}+(1/3) \left(C^{36}+2C^{45}\right),\\
{}^{\rm (tr)}P^{13}&=&C^{15}+C^{35}+(1/3)\left(C^{25}+2C^{46}\right),\\
{}^{\rm (tr)}P^{23}&=&C^{24}+C^{34}+(1/3)\left(C^{14}+2C^{56}\right)\,.
\end{eqnarray}
Using the 
$(3\times 7)$-matrix representation (\ref{S3}) we immediately derive 9  components of the matrix $R^{ij}$ 
\begin{eqnarray}
{}^{\rm (tr)}R^{11}&=&C^{11}- (6/7) P^{11}-(1/5) S,\\
{}^{\rm (tr)}R^{22}&=&C^{22}-(6/7) P^{22}-(1/5)  S,\\
{}^{\rm (tr)}R^{33}&=&C^{33}-(6/7) P^{33}-(1/5) S,\\
{}^{\rm (tr)}R^{12}&=&C^{16}-(3/7) P^{12},\qquad\qquad\quad {}^{\rm (tr)}R^{21}=C^{26}-(3/7) P^{12},\\
{}^{\rm (tr)}R^{13}&=&C^{15}-(3/7) P^{13},\qquad\qquad\quad {}^{\rm (tr)}R^{31}=C^{35}-(3/7) P^{13},\\
{}^{\rm (tr)}R^{23}&=&C^{24}-(3/7) P^{23},\qquad\qquad\quad {}^{\rm (tr)}R^{32}=C^{34}-(3/7) P^{23}.
\end{eqnarray}

The non-Cauchy part is represented by the  scalar $A$ and the symmetric traceless tensor $Q^{ij}$. From (\ref{sub8a}), we have the following expression for the scalar 
\begin{equation}\label{A-scalar}
{}^{\rm (tr)}A=(2/3)\left(C^{12}+C^{13}+C^{23}-C^{44}-C^{55}-C^{66}\right).
\end{equation}
The  components of the tensor $Q^{ij}$ are expressed as  
\begin{eqnarray}
{}^{\rm (tr)}Q^{11}&=&(2/3) \left(C^{23}-C^{44}\right)-(1/3) A,\qquad {}^{\rm (tr)}Q^{12}= C^{45}-C^{36}, \\
{}^{\rm (tr)}Q^{22}&=&(2/3) \left(C^{13}-C^{55}\right)-(1/3) A,\qquad  {}^{\rm (tr)}Q^{13}=  C^{46}-C^{25},\\
{}^{\rm (tr)}Q^{33}&=&(2/3) \left(C^{12}-C^{66}\right)-(1/3) A,\qquad {}^{\rm (tr)}Q^{23}= C^{56}-C^{14}.
\end{eqnarray}
Consequently, 21 elasticity modules of the triclinic system are described in our representation by their 21 independent  linear combinations  organized into two scalars and three matrices: 
\vspace{0.4cm}

{\centerline {\bf Triclinic system}}

\vspace{-0.5cm}

\begin{equation}\label{mono-9}\boxed{
S,\quad\!\!\!\! A,\quad \!\!\!\!
P^{ij}=\begin{bmatrix}
P^{11}&P^{12}&P^{13}\\ 
P^{12}&P^{22}&P^{23}\\ 
P^{13}&P^{23}&P^{33}
\end{bmatrix},\quad\!\!\!\!
Q^{ij}=\begin{bmatrix}
Q^{11}&Q^{12}&Q^{13}\\ 
Q^{12}&Q^{22}&Q^{23}\\ 
Q^{13}&Q^{23}&Q^{33}
\end{bmatrix},\quad\!\!\!\!
R^{ij}=\begin{bmatrix}
R^{11}&R^{12}&R^{13}\\ 
R^{21}&R^{22}&R^{23}\\ 
R^{31}&R^{32}&R^{33}
\end{bmatrix}}
\end{equation}
 
\subsection{Monoclinic system}
For a material with one mirror plane (and  a 2nd-order axis of rotation perpendicular to this plane), the number of independent components is reduced to 13. This fact follows from a simple symmetry argument:  Let the mirror plane be chosen as $x^3=0$. Then, under the reflection, all components of $C^{ijkl}$ that include index 3   one or three number of times will change sign while the other components will remain unchanged. The elasticity tensor must be preserved under  the reflection, thus all components with an odd number of index  3 must be zero. Consequently, we have eight {\it monoclinic constraints} 
\begin{equation}\label{mono-cond}
    C^{14}=C^{15}=C^{24}=C^{25}=C^{34}=C^{35}=C^{46}=C^{56}=0\,.
\end{equation}
Thus, we are left with $13=21-8$  independent  components that are are distributed in the body of  $(6\times 6)$-matrix as follows:
\begin{equation}\label{voigt-monocl}
{}^{\rm (mon)}C^{ijkl}= \begin{bmatrix}
  C^{11} & C^{12} & C^{13} & 0 & 0 & C^{16} \\
* & C^{22} & C^{23} & 0 & 0 & C^{26} \\
* & * & C^{33} & 0 & 0 & C^{36} \\
* & * & * & C^{44} & C^{45} & 0 \\
* & * & * & * & C^{55} & 0 \\
* & * & * & * & * & C^{66}\, \end{bmatrix}.
\end{equation}
In $(3\times 7)$-matrix notation, the monoclinic elasticity tensor is decomposed into a sum of  two independent pieces with the dimensions  $9\oplus 4=13$ as follows: 
\begin{eqnarray}\label{new1a}
{}^{\rm (mon)}C^{ijkl}&=&
\begin{bmatrix}
C^{11}&C^{22}&C^{33}\\ 
C^{16}&C^{26}&0\\ 
0&0&0\\ 
C^{12}&C^{13}&C^{23}\\
C^{66}&C^{55}&C^{44}\\ 
0&0&C^{36}\\ 
0&0&C^{45}
\end{bmatrix}
=
\frac13\begin{bmatrix}
3\pmb {C^{11}}&3\pmb {C^{22}}&3\pmb {C^{33}}\\ 
3\pmb {C^{16}}&3\pmb {C^{26}}&0\\ 
0&0&0\\ 
\pmb {C^{12}+2C^{66}}&\pmb{C^{13}+2 C^{55}}&\pmb {C^{23}+2C^{44}}\\
C^{12}+2C^{66}&C^{13}+2C^{55}&C^{23}+2C^{44}\\
0&0&\pmb {C^{36}+2C^{45}}\\
0&0&C^{36}+2C^{45}
\end{bmatrix}
+\nonumber\\
&&\frac13\begin{bmatrix}
0&0&0\\ 
0&0&0\\ 
0&0&0\\ 
2\pmb {\left(C^{12}-C^{66}\right)}&2\pmb {\left(C^{13}-C^{55}\right)}&2\pmb {\left(C^{23}-C^{44}\right)}\\
-\left(C^{12}-C^{66}\right)&-\left(C^{13}-C^{55}\right)&-\left(C^{23}-C^{44}\right)\\ 
0&0&2\pmb {\left(C^{36}-C^{45}\right)}\\
0&0&-\left(C^{36}-C^{45}\right)
\end{bmatrix}.
\end{eqnarray}
Recall that we use the bold font  to visualize the independent components. As it pointed out in \cite{Landau}, only the $z$-axis  is fixed in (\ref{voigt-monocl}) and in (\ref{new1a}). With an appropriate rotation of the $(x,y)$ coordinate system, the term $C^{1233}\equiv C^{36}$ can be removed. In fact, with the same rotation we can remove the term $C^{36}+2C^{45}\equiv  C^{1233}+2C^{1323}$ in the first piece or the term  $C^{36}-C^{45}\equiv C^{1233}-C^{1323}$ in the second piece of  (\ref{new1a}). 

We present now the $SO(3)$ decomposition of  two tensors given in (\ref{new1a}). It is enough to substitute the monoclinic conditions (\ref{mono-cond}) into the corresponding triclinic terms. Two scalars for the monoclinic system   are left exactly the same as for the triclinic system
\begin{equation}\label{mono-scalars} 
    {}^{\rm (mon)}S={}^{\rm (tr)}S\,,\qquad
      {}^{\rm (mon)}A={}^{\rm (tr)}A\,.
\end{equation}
The second-order tensor parts are determined by the matrices $P^{ij}$ and $Q^{ij}$ with  the  additional  constraints 
\begin{equation}\label{mono-ten1}
{}^{\rm (mon)}P^{13}={}^{\rm (mon)}P^{23}=0\,,\qquad
{}^{\rm (mon)}Q^{13}={}^{\rm (mon)}Q^{23}=0\,.
\end{equation}
As for the nonzero components of these tensors, they are left  exactly the same as for the triclinic material, i.e., for the pairs $\{ij\}\ne \{13\}$ and $\{ij\}\ne \{23\}$, we have
\begin{equation}\label{mono-ten2}
{}^{\rm (mon)}P^{ij}={}^{\rm (tr)}P^{ij}=0\,,\qquad
{}^{\rm (mon)}Q^{ij}={}^{\rm (tr)}Q^{ij}=0\,.
\end{equation}
The third piece of the Cauchy part is completely characterized by the matrix $R^{ij}$ with the constraints 
\begin{equation}\label{mono-R1}
{}^{\rm (mon)}R^{13}={}^{\rm (mon)}R^{23}={}^{\rm (mon)}R^{31}={}^{\rm (mon)}R^{32}=0\,.
\end{equation}
Thus $R^{ij}$ is left with 5  independent  components that have exactly the same expressions as for the triclinic system 
\begin{equation}\label{mono-R2}
{}^{\rm (mon)}R^{ij}={}^{\rm (tr)}R^{ij}=0\,, \qquad \mbox{for} \quad i,j\ne 3\,.
\end{equation}
Notice that instead of  direct substitution of eight relations (\ref{mono-cond}) into the triclinic expressions, the terms  (\ref{mono-ten1}) and (\ref{mono-R1}) can be derived straightforwardly by a simple  symmetry argument: these expressions vanish because they have an odd number of the index 3. 

Consequently, the monoclinic system is described by two scalars $S$ and $A$ and three matrices of the following form 
\vspace{0.4cm}

{\centerline {\bf Monoclinic system}}

\vspace{-0.5cm}

\begin{equation}\label{mono-9}
\boxed{S,\quad\!\!\!\! A,\quad\!\!\!\!
P^{ij}=\begin{bmatrix}
P^{11}&P^{12}&0\\ 
P^{12}&P^{22}&0\\ 
0&0&P^{33}
\end{bmatrix},\quad\!\!\!\!
Q^{ij}=\begin{bmatrix}
Q^{11}&Q^{12}&0\\ 
Q^{12}&Q^{22}&0\\ 
0&0&Q^{33}
\end{bmatrix},\quad\!\!\!\!
R^{ij}=\begin{bmatrix}
R^{11}&R^{12}&0\\ 
R^{21}&R^{22}&0\\ 
0&0&R^{33}
\end{bmatrix}}
\end{equation}
As it was mentioned above,  we can remove the non-diagonal terms $P^{12}$ or $Q^{12}$ with an appropriate rotation of the coordinate system $(x,y)$. It cannot be done, however, for the terms $R^{12}$ or $R^{21}$ because the matrix $R^{ij}$ does not  a tensor and its components do not transform by an ordinary transformation law.  

\subsection{Orthotropic (rhombic) system}
These materials have three mutually orthonormal mirror planes. They also can be characterized by three mutually perpendicular twofold axes. 
 When the planes are chosen to be the basic planes of  Cartesian system, the only nonzero components of $C^{ijkl}$ are those whose indices contain each one of values 1,2, or 3  even number of times. 
Thus, in  addition to eight monoclinic conditions (\ref{mono-cond}), we have four pure  orthotropic conditions
 \begin{equation}\label{rhomb-cond}
 C^{16}=C^{26}=C^{36}=C^{45}=0\,.
 \end{equation}
 Consequently,  the tensor  $C^{ijkl}$ of the  orthotropic  system has only $9=13-4$ independent components. In Voigt's $(6\times 6)$-matrix notation, it reads 
\begin{equation}\label{rhomb1}
{}^{\rm{(rh)}}C^{ijkl}=\begin{bmatrix}
  C^{11} & C^{12} & C^{13} & 0 & 0 & 0 \\
* & C^{22} & C^{23} & 0 & 0 & 0 \\
* & * & C^{33} & 0 & 0 & 0 \\
* & * & * & C^{44} & 0 & 0 \\
* & * & * & * & C^{55} & 0 \\
* & * & * & * & * & C^{66}\, \end{bmatrix}.
\end{equation}
The elasticity tensor is decomposed to the Cauchy and non-Cauchy parts with the dimensions  $9=6\oplus 3$. In $(3\times 7)$-matrix form, we have 
\begin{eqnarray}\label{new1a}
{}^{\rm{(rh)}}C^{ijkl}&=&
\begin{bmatrix}
C^{11}&C^{22}&C^{33}\\ 
0&0&0\\ 
0&0&0\\ 
C^{12}&C^{13}&C^{23}\\
C^{66}&C^{55}&C^{44}\\ 
0&0&0\\ 
0&0&0
\end{bmatrix}
=
\frac13\begin{bmatrix}
3\pmb {C^{11}}&3\pmb {C^{22}}&3\pmb {C^{33}}\\ 
0&0&0\\ 
0&0&0\\ 
\pmb {C^{12}+2C^{66}}&\pmb{C^{13}+2 C^{55}}&\pmb {C^{23}+2C^{44}}\\
C^{12}+2C^{66}&C^{13}+2C^{55}&C^{23}+2C^{44}\\
0&0&0\\
0&0&0
\end{bmatrix}
+\nonumber\\
&&\frac13\begin{bmatrix}
0&0&0\\ 
0&0&0\\ 
0&0&0\\ 
2\pmb {\left(C^{12}-C^{66}\right)}&2\pmb {\left(C^{13}-C^{55}\right)}&2\pmb {\left(C^{23}-C^{44}\right)}\\
-\left(C^{12}-C^{66}\right)&-\left(C^{13}-C^{55}\right)&-\left(C^{23}-C^{44}\right)\\ 
0&0&0\\
0&0&0
\end{bmatrix}.
\end{eqnarray}

We turn now to the $SO(3)$-decomposition. 
The scalar parts of the elasticity tensor are exactly the same as for monoclinic (and triclinic) system 
\begin{equation}\label{mono-scalar}
    {}^{\rm{(rh)}}S={}^{\rm{(mon)}}S\,,\qquad 
    {}^{\rm{(rh)}}A={}^{\rm{(mon)}}A\,.
\end{equation}
For the tensors $P^{ij}$ and $Q^{ij}$, the components with $i\ne j$ change their sign under reflections. Thus they must be equal zero. Therefore, we have two traceless diagonal matrices with two independent components each. The same symmetry argument is applicable  also to the matrix $R^{ij}\equiv R^{iiij}$, that represents the components of the tensor with three equal indices. Hence, we are left with only 3 diagonal  components of this matrix, $R^{11},R^{22},R^{33}$. 

Consequently, orthotropic  system is described by two scalars $S$ and $A$ and three diagonal matrices 
\vspace{0.4cm}

{\centerline {\bf Orthotropic  system}}

\vspace{-0.5cm}

\begin{equation}\label{S2-ortho}
\boxed{S,\quad\!\!\!\! A,\quad\!\!\!\!
P^{ij}=\begin{bmatrix}
P^{11}&0&0\\
0&P^{22}&0\\
0&0&P^{33}
\end{bmatrix}\,, \quad\!\!\!\! Q^{ij}=\begin{bmatrix}
Q^{11}&0&0\\
0&Q^{22}&0\\
0&0&Q^{33}
\end{bmatrix}\,, 
 \quad\!\!\!\!
R^{ij}=\begin{bmatrix}
R^{11}&0&0\\
0&R^{22}&0\\
0&0&R^{33}
\end{bmatrix}}
\end{equation}
Notice that all elements presented here   are expressed by the  elasticity modules $C^{IJ}$ exactly as the corresponding components of the monoclinic and triclinic systems. 
\subsection{Trigonal system}
This system is characterized by one  three-fold axis of rotation. As it is shown in \cite{Landau} and \cite{Love},  there are  7 independent modules, that enter  Voigt's $(6\times 6)$-matrix as 
\begin{equation}\label{trig1-0}
{}^{\rm{(tr)}}C^{ijkl}= \begin{bmatrix}
  C^{11} & C^{12} & C^{13} & C^{14} & C^{15} & 0 \\
* & C^{11} & C^{13} &- C^{14} & -C^{15} & 0 \\
* & * & C^{33} & 0 & 0 & 0 \\
* & * & * & C^{44} & 0 & -C^{15} \\
* & * & * & * & C^{44} & C^{14} \\
* & * & * & * & * & C^{66}\, \end{bmatrix}
\end{equation}
with an additional condition 
\begin{equation}\label{trig1-1}
C^{66}=(1/2) (C^{11}-C^{12}).
\end{equation}
The decomposition of this tensor into the Cauchy and non-Cauchy irreducible parts splits the dimension as $7=5\oplus 2$. In $(3\times 7)$-matrix notation,  
\begin{eqnarray}\label{trig1-2}
{}^{\rm (tr)}C^{ijkl}&=&
\begin{bmatrix}
C^{11}&C^{11}&C^{33}\\ 
0&0&0\\ 
C^{15}&-C^{14}&0\\ 
C^{12}&C^{13}&C^{13}\\
C^{66}&C^{44}&C^{44}\\ 
C^{14}&-C^{15}&0\\ 
C^{14}&-C^{15}&0
\end{bmatrix}=
\frac13\begin{bmatrix}
3\pmb {C^{11}}&3 {C^{11}}&3\pmb {C^{33}}\\
0&0&0\\ 
3\pmb {C^{15}}&-3\pmb {C^{14}}&0\\ 
 {C^{11}}&\pmb{C^{13}+2 C^{44}}& {C^{13}+2C^{44}}\\
C^{11}&C^{13}+2C^{44}&C^{13}+2C^{44}\\
 {3C^{14}}& {-3C^{15}}& 0\\
3C^{14}&-3C^{15}&0
\end{bmatrix}\nonumber\\&&
+\frac13\begin{bmatrix}
0&0&0\\ 
0&0&0\\ 
0&0&0\\ 
2\pmb {\left(C^{12}-C^{66}\right)}&2\pmb {\left(C^{13}-C^{44}\right)}&2 {\left(C^{13}-C^{44}\right)}\\
-\left(C^{12}-C^{66}\right)&-\left(C^{13}-C^{44}\right)&-\left(C^{13}-C^{44}\right)\\ 
0&0&0\\
0&0&0
\end{bmatrix}.
\end{eqnarray}
To provide the $SO(3)$ decomposition, we consider first the Cauchy part. The scalar invariant takes the form
\begin{equation}\label{trig1-3}
S=(4/3)(2C^{11}+2C^{44}+C^{13})+C^{33}\,.
\end{equation}
The second-order tensor piece $P_{ij}$ is described by a diagonal scalar  matrix with only one independent component. We define 
\begin{equation}\label{trig1-4}
P:=P^{11}=P^{22}=-(1/2) P^{33}. 
\end{equation}
Then, by comparison of (\ref{trig1-2}) with (\ref{S2a}), we have 
\begin{equation}\label{trig1-4a}
P=(1/3) (4C^{11}+2C^{44}+C^{13}-S)\,.
\end{equation}
The third part is described by the matrix $R^{ij}$. We denote  its diagonal elements  
\begin{equation}\label{trig1-5}
R:=R^{11}=R^{22}=(3/8) R^{33},
\end{equation}
and derive from (\ref{S2a})
\begin{equation}\label{trig1-5a}
R=C^{11}-(6/7)P^{11}-(1/5)S\,.
\end{equation}
In addition we have in $R^{ij}$  two non-diagonal elements
\begin{equation}\label{trig1-6}
R^{13}=C^{15}\,,\qquad R^{23}=-C^{14}\,.
\end{equation}

The non-Cauchy part of the trigonal elasticity tensor has 2 independent components. The scalar invariant is expressed as 
\begin{equation}\label{trig1-7}
A=(2/3)(C^{11}+2C^{13}-2C^{44}-3C^{66})\,.
\end{equation}
The  matrix $Q^{ij}$ is scalar and  diagonal. We denote 
\begin{equation}\label{trig1-8}
Q:=Q^{11}=Q^{22}=-(1/2) Q^{33},
\end{equation}
then we have
\begin{equation}\label{trig1-8a}
Q=(1/3) (2C^{13}-2C^{44}-A).
\end{equation}
Consequently, this type of material is completely  characterized by two scalars $S$ and $A$, two scalar tensors $P^{ij}$ and $Q^{ij}$ and the matrix $R^{ij}$ of 3 independent components.
\vspace{0.4cm}

{\centerline {\bf Trigonal(7)  system}}

\vspace{-0.5cm}

\begin{equation}\label{trig1-9}
\boxed{S,\quad A,\quad
P^{ij}=P\begin{bmatrix}
1&0&0\\ 
0&1&0\\ 
0&0&-2
\end{bmatrix},\quad
Q^{ij}=Q\begin{bmatrix}
1&0&0\\ 
0&1&0\\ 
0&0&-2
\end{bmatrix},\quad
R^{ij}=\begin{bmatrix}
R&0&R^{13}\\ 
0&R&R^{23}\\ 
0&0&(8/3)R
\end{bmatrix}}
\end{equation}
Notice that this presentation of the trigonal system by two scalars, $S$ and $A$, and three simple matrices (\ref{trig1-9}) is much more illustrative than the ordinary presentation (\ref{trig1-0}) with the additional requirement (\ref{trig1-1}). 

It is well known, see e.g. \cite{Cowin1995b}, \cite{Moakher}, that with an appropriate rotation of the coordinate system around $z$-axis one  elasticity module  can be vanished, in particular $C^{14}=0$. The  elasticity tensor of a such reduced trigonal system has only 6 independent modules 
\begin{equation}\label{trig2-1}
{}^{\rm{(tr)}}C^{ijkl}=\begin{bmatrix}
  C^{11} & C^{12} & C^{13} & 0 & C^{15} & 0 \\
* & C^{11} & C^{13} &0 & -C^{15} & 0 \\
* & * & C^{33} & 0 & 0 & 0 \\
* & * & * & C^{44} & 0 & -C^{15} \\
* & * & * & * & C^{55} & 0 \\
* & * & * & * & * & C^{66}\, \end{bmatrix}.
\end{equation}
with
\begin{equation}\label{trig2-2}C^{66}=\frac 12 (C^{11}-C^{12})\end{equation}
It is interesting to look which part of the elasticity tensor is sensitive to this reduction. 
As a subset of the previous case, the elasticity tensor (\ref{trig2-1}) is characterized by 
two scalars $S$ and $A$, two scalar tensors $P^{ij}$ and $Q^{ij}$ and the matrix $R^{ij}$ of 2 independent entries. Consequently, the reduced trigonal system is represented as
\vspace{0.4cm}

{\centerline {\bf Trigonal(6)  system}}

\vspace{-0.5cm}

\begin{equation}\label{trig2-9}
\boxed{S,\quad A,\quad
P^{ij}=P\begin{bmatrix}
1&0&0\\ 
0&1&0\\ 
0&0&-2
\end{bmatrix},\quad
Q^{ij}=Q\begin{bmatrix}
1&0&0\\ 
0&1&0\\ 
0&0&-2
\end{bmatrix},\quad
R^{ij}=\begin{bmatrix}
R&0&R^{13}\\ 
0&R&0\\ 
0&0&(8/3)R
\end{bmatrix}}
\end{equation}
The expressions of the scalars $S$ and $A$, the coefficients $P$ and $Q$, and of the components $R$ and $R^{13}$ via the modules $C^{ij}$ are left exactly the same as given in the previous case. Only the element $R^{23}$ vanishes in this special coordinate system, see (\ref{trig1-6}).
\subsection{Transverse isotropy (hexagonal)   system}
These crystal system is characterized by a sixth-order axis of rotation. In the plane normal to the rotation axis, the deformation properties of the material are the same as for an isotropic body -- transverse isotropy.
There are 5 independent constants, see \cite{Landau} 
\begin{equation}\label{hex1}
C^{ijkl}= \begin{bmatrix}
  C^{11} & C^{12} & C^{13} & 0 & 0 & 0 \\
* & C^{11} & C^{13} &0 &0 & 0 \\
* & * & C^{33} & 0 & 0 & 0 \\
* & * & * & C^{44} & 0 & 0 \\
* & * & * & * & C^{44} & 0 \\
* & * & * & * & * & C^{66}\, \end{bmatrix}.
\end{equation}
$$C^{66}=\frac 12 (C^{11}-C^{12})$$
This case can be considered as a subset of the reduced trigonal system with a successive  restriction $C^{15}=0$. In this case,  $R^{13}=0$ and the matrix $R^{ij}$ turns out to be diagonal and even scalar. 
Consequently, the hexagonal  materials 
are characterized by the same two scalar $S$ and $A$ as in the trigonal system (\ref{trig1-3}) and (\ref{trig1-7}), respectively, and three scalar matrices
\vspace{0.4cm}

{\centerline {\bf Hexagonal  system}}

\vspace{-0.5cm}

\begin{equation}\label{hex2}
\boxed{S,\quad A,\quad
P^{ij}=P\begin{bmatrix}
1&0&0\\ 
0&1&0\\ 
0&0&-2
\end{bmatrix},\quad
Q^{ij}=Q\begin{bmatrix}
1&0&0\\ 
0&1&0\\ 
0&0&-2
\end{bmatrix},\quad
R^{ij}=R\begin{bmatrix}
1&0&0\\ 
0&1&0\\ 
0&0&(8/3)
\end{bmatrix}.}
\end{equation}
The leading coefficients here are the same as for the trigonal system
\begin{eqnarray}
P&=&(1/3) (4C^{11}+2C^{44}+C^{13}-S),\\
Q&=&(1/3) (2C^{13}-2C^{44}-A),\\
R&=&C^{11}-(6/7)P-(1/5)S
\end{eqnarray}

\subsection{Tetragonal  system}
These materials have a fourth-order axis of rotation. 
When the axis is taken as the $z$ axis of the coordinate system, the elasticity tensor includes  7 independent components, see \cite{Landau}. In Voigt's notations, it reads
\begin{equation}\label{tet1-1}
{}^{\rm (tet)}C^{ijkl}=\begin{bmatrix}
  C^{11} & C^{12} & C^{13} & 0 & 0 & C^{16} \\
* & C^{11} & C^{13} &0 &0 & -C^{16} \\
* & * & C^{33} & 0 & 0 & 0 \\
* & * & * & C^{44} & 0 & 0 \\
* & * & * & * & C^{44} & 0 \\
* & * & * & * & * & C^{66}\, \end{bmatrix}.
\end{equation}
In $(3\times 7)$-matrix notation, the elasticity tensor and its Cauchy and non-Cauchy parts are represented, respectively, as 
\begin{eqnarray}\label{tet1-2}
{}^{\rm (tet)}C^{ijkl}&=&
\begin{bmatrix}
C^{11}&C^{11}&C^{33}\\ 
C^{16}&-C^{16}&0\\ 
0&0&0\\ 
C^{12}&C^{13}&C^{13}\\
C^{66}&C^{44}&C^{44}\\ 
0&0&0\\ 
0&0&0
\end{bmatrix}
=
\frac 13\begin{bmatrix}
3\pmb {C^{11}}&3C^{11}&3\pmb {C^{33}}\\ 
3\pmb {C^{16}}&-3C^{16}&0\\ 
0&0&0\\ 
\pmb {C^{12}+2C^{66}}&\pmb {C^{13}+2C^{44}}&C^{13}+2C^{44}\\
C^{12}+2C^{66}&C^{13}+2C^{44}&C^{13}+2C^{44}\\
0&0&0\\ 
0&0&0
\end{bmatrix}
+\nonumber\\
&&\frac 13\begin{bmatrix}
0&0&0\\ 
0&0&0\\ 
0&0&0\\ 
2\pmb {(C^{12}-C^{66})}&2\pmb {(C^{13}-C^{44})}&2(C^{13}-C^{44})\\
-(C^{12}-C^{66})&-(C^{13}-C^{44})&-(C^{13}-C^{44})\\
0&0&0\\ 
0&0&0
\end{bmatrix}.
\end{eqnarray}
The tensor space of the elasticity tensor is decomposed into the direct sum of two subspace with the dimensions $7=5\oplus 2$.

We start the $SO(3)$-decomposition with the Cauchy part. The scalar $S$ takes the form  
\begin{equation}\label{tet1-3}
{}^{\rm (tet)}S=2C^{11}+C^{33}+(2/3)(C^{12}+2C^{13}+4C^{44}+2C^{66})\,.
\end{equation}
The non-diagonal elements of the matrix $P^{ij}$ vanish. Denote  
\begin{equation}\label{tet1-4}
P:=P^{11}=P^{22}=-(1/2)P^{33}\,.
\end{equation}
Then we derive 
\begin{equation}\label{tet1-4a}
P=C^{11}
+(1/3)(C^{12}+C^{13}+2C^{44}+2C^{66}-S)\,.
\end{equation}
The $R$-matrix is 3-parametric and antisymmetric. Its independent components are expressed as 
\begin{eqnarray}\label{tet1-6}
R^{11}&=&R^{22}=C^{11}-(6/7)P-(1/5)S\,,\\
R^{33}&=&C^{33}+(12/7)P-(1/5)S\,,\\
R^{12}&=&-R^{21}=C^{16}.
\end{eqnarray}

The non-Cauchy part is 2-dimensional and described by a scalar
\begin{equation}\label{tet1-8}
{}^{\rm (tet7)}A=(2/3)(2C^{12}-C^{13}-2C^{44}-C^{66})\,.
\end{equation}
and a scalar matrix $Q^{ij}$.  With 
\begin{equation}\label{tet1-9}
Q:=Q^{11}=Q^{22}=-(1/2)Q^{33},
\end{equation}
we derive
\begin{equation}\label{tet1-9a}
Q=(1/3)(2C^{13}-2C^{44}-A)\,.
\end{equation}
Therefore, the tetragonal system is represented by two scalars $S$ and $A$, and two scalar matrices $P^{ij}$ and  $Q^{ij}$, and a  matrix $R^{ij}$ of three independent components:

\vspace{0.4cm}

{\centerline {\bf Tetragonal(7)  system}}

\vspace{-0.5cm}

\begin{equation}\label{tet1-5a}
\boxed{S,\quad\!\!\! A,\quad\!\!\!
P^{ij}=P\begin{bmatrix}
1&0&0\\ 
0&1&0\\ 
0&0&-2
\end{bmatrix},
\quad \!\!\!
Q^{ij}=Q\begin{bmatrix}
1&0&0\\ 
0&1&0\\ 
0&0&-2
\end{bmatrix},
\quad \!\!\!
R^{ij}=\begin{bmatrix}
R^{11}&R^{12}&0\\ 
-R^{12}&R^{11}&0\\ 
0&0&R^{33}
\end{bmatrix}}
\end{equation}

With a special rotation of the coordinate system, the module $C^{16}$  can be removed.   This way, we obtain a reduced tetragonal system with 6 independent components. 
This case can be considered as a subset of the previous one with an additional condition $C^{16}=0$.  Notice that this value does  not enter the scalars $S$ and $A$ and not the tensors $P^{ij}$ and $Q^{ij}$. Only for the matrix $R^{ij}$ it yields  $R^{12}=0$. Thus $R^{ij}$  turns out to be diagonal  with the same values on the diagonal as in the previous case. 
Consequently, the reduced tetragonal system is described with two scalars $S$ and $A$, and three diagonal matrices
\vspace{0.4cm}

{\centerline {\bf  Tetragonal(6)  system}}

\vspace{-0.5cm}

\begin{equation}\label{tet1-5ab}
\boxed{S,\quad\!\!\! A,\quad\!\!\!
P^{ij}=P\begin{bmatrix}
1&0&0\\ 
0&1&0\\ 
0&0&-2
\end{bmatrix},
\qquad \!\!\!
Q^{ij}=Q\begin{bmatrix}
1&0&0\\ 
0&1&0\\ 
0&0&-2
\end{bmatrix},
\qquad \!\!\!
R^{ij}=\begin{bmatrix}
R^{11}&0&0\\ 
0&R^{11}&0\\ 
0&0&R^{33}
\end{bmatrix}}
\end{equation}

\subsection{Cubic  system}
Cubic system represents a simplest anisotropic material. 
There are 3 independent elasticity module, that are arranged in the $6\times 6$ matrix as 
\begin{equation}\label{cub0}
{}^{\rm (cub)}C^{ijkl}= \begin{bmatrix}
  C^{11} & C^{12} & C^{12} & 0 & 0 & 0 \\
* & C^{11} & C^{12} &0 &0 & 0 \\
* & * & C^{11} & 0 & 0 & 0 \\
* & * & * & C^{44} & 0 & 0 \\
* & * & * & * & C^{44} & 0 \\
* & * & * & * & * & C^{44}\, \end{bmatrix}.
\end{equation}
Cubic system can be considered as a subset of the tetragonal  one with a set of  additional conditions \begin{equation}\label{cub-1}
C^{13}=C^{12},\quad C^{33}=C^{11},\qquad C^{66}=C^{44}.
\end{equation}
In $3\times 7$ form, the elasticity tensor is decomposed to the Cauchy and non-Cauchy parts as $3=2\oplus 1$. Explicitly we have 
\begin{eqnarray}\label{cub-2}
{}^{\rm (cub)}C^{ijkl}&=&\begin{bmatrix}
C^{11}&C^{11}&C^{11}\\ 
0&0&0\\ 
0&0&0\\ 
C^{12}&C^{12}&C^{12}\\
C^{44}&C^{44}&C^{44}\\ 
0&0&0\\ 
0&0&0
\end{bmatrix}
=
\frac 13\begin{bmatrix}
3\pmb{C^{11}}&3C^{11}&3C^{11}\\ 
0&0&0\\ 
0&0&0\\ 
\pmb{C^{12}+2C^{44}}&C^{12}+2C^{44}&C^{12}+2C^{44}\\
C^{12}+2C^{44}&C^{12}+2C^{44}&C^{12}+2C^{44}\\
0&0&0\\ 
0&0&0
\end{bmatrix}
+\nonumber\\
&&\frac 13\begin{bmatrix}
0&0&0\\ 
0&0&0\\ 
0&0&0\\ 
2\pmb{(C^{12}-C^{44})}&2(C^{12}-C^{44})&2(C^{12}-C^{44})\\
-(C^{12}-C^{44})&-(C^{12}-C^{44})&-(C^{12}-C^{44})\\
0&0&0\\ 
0&0&0
\end{bmatrix}.
\end{eqnarray}
The Cauchy and non-Cauchy scalars, respectively,  take the form  
\begin{equation}\label{cub-3}
{}^{\rm (cub)}S=3C^{11}+2C^{12}+4C^{44}\,.
\end{equation}
and 
\begin{equation}\label{cub-4}
{}^{\rm (cub)}A=(2/3)(C^{12}-3C^{44})\,.
\end{equation}
From (\ref{tet1-4a}) we obtain that the tensors $P^{ij}$ and $Q^{ij}$ are identically zero. From (\ref{tet1-6}) it follows that the matrix $R^{ij}$ is proportional to the unit matrix $R^{ij}=R\,{\rm diag} (1,1,1)$ with
\begin{equation}\label{tet1-6}
R=(2/5)(C^{11}-C^{12}-2C^{44})\,.
\end{equation}
Consequently the cubic system is described by two scalars and one scalar matrix: 

\vspace{0.4cm}

{\centerline {\bf   Cubic  system}}


\begin{equation}\label{cub10}
\boxed{S,\quad A,\quad P^{ij}=0,\quad Q^{ij}=0,\quad 
R^{ij}=R\begin{bmatrix}
1&0&0\\ 
0&1&0\\ 
0&0&1
\end{bmatrix}.}
\end{equation}

\subsection{Isotropic materials}
Isotropic material can be considered as a subset of the cubic one with an additional requirement 
\begin{equation}\label{iso-1}
C^{12}=C^{11}-2C^{44}\,.
\end{equation}
Consequently, three matrices vanish
\begin{equation}\label{iso-2}
P^{ij}=Q^{ij}=R^{ij}=0\,,
\end{equation}
while the scalars are reduced to the form
\begin{equation}\label{iso-3}
{}^{\rm (iso)}S=5(C^{12}+2C^{44})=5(\lambda+2\mu)
\end{equation}
and 
\begin{equation}\label{cub-4}
{}^{\rm (iso)}A=(2/3)(C^{12}-3C^{44})=(2/3)(\lambda-3\mu)\,.
\end{equation}
Here we use the standard Lame modules  $C^{12}=\lambda$ and $C^{44}=\mu$ are used. Consequently, 

\vspace{0.4cm}

{\centerline {\bf  Isotropic  system}}
\begin{equation}\label{iso10}
\boxed{S,\quad A,\quad P^{ij}=0,\quad Q^{ij}=0,\quad 
R^{ij}=0.}
\end{equation}

\section{Symmetry systems hierarchy}
The hierarchy of the crystal symmetry systems is an important issue for a lot of subjects in elasticity, in particular, for the problem of  averaging the elasticity tensor of a low-symmetry crystal by a higher symmetry prototype -- generalized Fedorov problem \cite{Moakher}, see also \cite{Weber} for recent study. 
Different non-equivalent hierarchy diagrams often appear in elasticity and acoustic literature, see for instance \cite{Gazis}, \cite{Moakher}, \cite{Bona}. To our knowledge, there is not yet a generally accepted agreement on this subject.  In this section, we present a hierarchy  diagram based on the irreducible content of the symmetry systems. 
Usually such diagrams are constructed by embedding  of the full elasticity tensor. 
Our classification is based on a stronger requirement for  every irreducible part of a higher symmetric system to be properly embedded into the corresponding irreducible part of the lower symmetry system. 

In Tab. \ref{tab.1}, we collect  our results concerning the dimensions of the subspaces of elasticity tensor for different symmetry  systems. Recall that this decomposition is irreducible and unique. 
\begin{table}[!htb]
    \centering
    \begin{tabular}{clcccccc}
      Notation&Symmetry system  &$C^{ijkl}$& ${}^{(1)}S^{ijkl}$ & ${}^{(2)}S^{ijkl}$ & ${}^{(3)}S^{ijkl}$ &${}^{(1)}A^{ijkl}$&${}^{(2)}A^{ijkl}$\\
      \hline 
      A&triclinic &21 & 1 & 5 & 9 & 1 & 5\\
      B&monoclinic &13 & 1 & 3 & 5 & 1 & 3\\
      C&orthotropic  &9 & 1 & 2 & 3 & 1 & 2\\
      D&trigonal-7 &7 & 1 & 1 & 3& 1 & 1\\
      E&trigonal-6 &6 & 1 & 1 & 2& 1 & 1\\
      F&tetragonal-7 &7 & 1 & 1 & 3& 1 & 1\\
      G&tetragonal-6 &6 & 1 & 1 & 2& 1 & 1\\
      H& transverse isotropy &5 & 1 & 1 & 1& 1 & 1\\
      I&cubic &3 & 1 & 0 & 1 & 1 & 0\\
      J&isotropic &2 & 1 & 0 & 0 & 1 & 0\\
    \end{tabular}
\caption{Irreducible parts of the elasticity tensor with their dimensions for all symmetry systems. }\label{tab.1}
\end{table}

First we observe that the one dimensional scalar parts ${}^{(1)}S^{ijkl}$ and ${}^{(1)}A^{ijkl}$ are included in all symmetry systems. Thus they are irrelevant for the classification problem. The  tensor parts ${}^{(2)}S^{ijkl}$ and ${}^{(2)}A^{ijkl}$ 
are completely described by two second-order tensors $P^{ij}$ and $Q^{ij}$. For every symmetry class, the dimensions of the $P$-spaces and the $Q$-spaces are the same. 
In other words,  the symmetry group of a  crystal cannot distinguish between  the tensors $P^{ij}$ and $Q^{ij}$.  
For most systems, these tensors are presented by the same one-parametric (scalar) matrices.  So they do not  enough for properly classification. 
The main difference between the symmetry systems  appears in the fourth-order tensor part ${}^{(3)}S^{ijkl}$, that is represented in our approach by the matrix $R^{ij}$. 
Although this matrix depends of the choice of coordinates, it is  applicable for the classification problem since we use in all crystal systems the same coordinate frame with the $z$-axis directed as the rotational axis. 

Comparing the matrices $P,Q$ and $R$ we see that the monoclinic system is properly embedded in the triclinic one. Similarly all higher symmetry systems, but the trigonal  one, are embedded into the monoclinic system. 

As for the trigonal type systems, their $R$ matrix (\ref{trig1-9}) and (\ref{trig2-9}) cannot be considered as a special case of the monoclinic $R$ matrix  (\ref{mono-9}).  Thus trigonal  system must be treated as a separate branch outgoing  from the triclinic system. 
This trigonal branch goes directly to the hexagonal system and then to the isotropic one, where the $R$-matrix vanishes. 
Comparison between the $P,Q$ and $R$-matrices of the orthotropic and non-reduced tetragonal system shows that they cannot be considered as the subsets of one other and must be viewed as two separate branches. This problem is immediately solved, however, when we turn to the reduced tetragonal system that turns to be a proper subset of the orthotropic one. The last issue  to consider is the relation between the cubic and the hexagonal (transverse isotropic) systems. The structure of their $R$-matrices shows that they are not subsets of one another and thus must be considered as two separate  branches. 
All other inclusions are obvious so we came the diagram depicted at Fig. 1.

This result is in a correspondence with the diagrams given in \cite{Gazis} and in \cite{Moakher}. Notice that in \cite{Moakher}, but not in  \cite{Gazis}, there is an additional inclusion of the cubic system into the non-reduced trigonal system. This pass is forbidden in our approach because of the different structures of the $R$-matrices. 
Our diagram  is different from the schemes given in  \cite{Bona}, \cite{}, and \cite{Weber} where the  trigonal system is considered as a sub-family of the monoclinic one.    

\begin{figure}[ph]\label{fig1}
\vspace*{-1.5cm}
    \centerline{\psfig{file=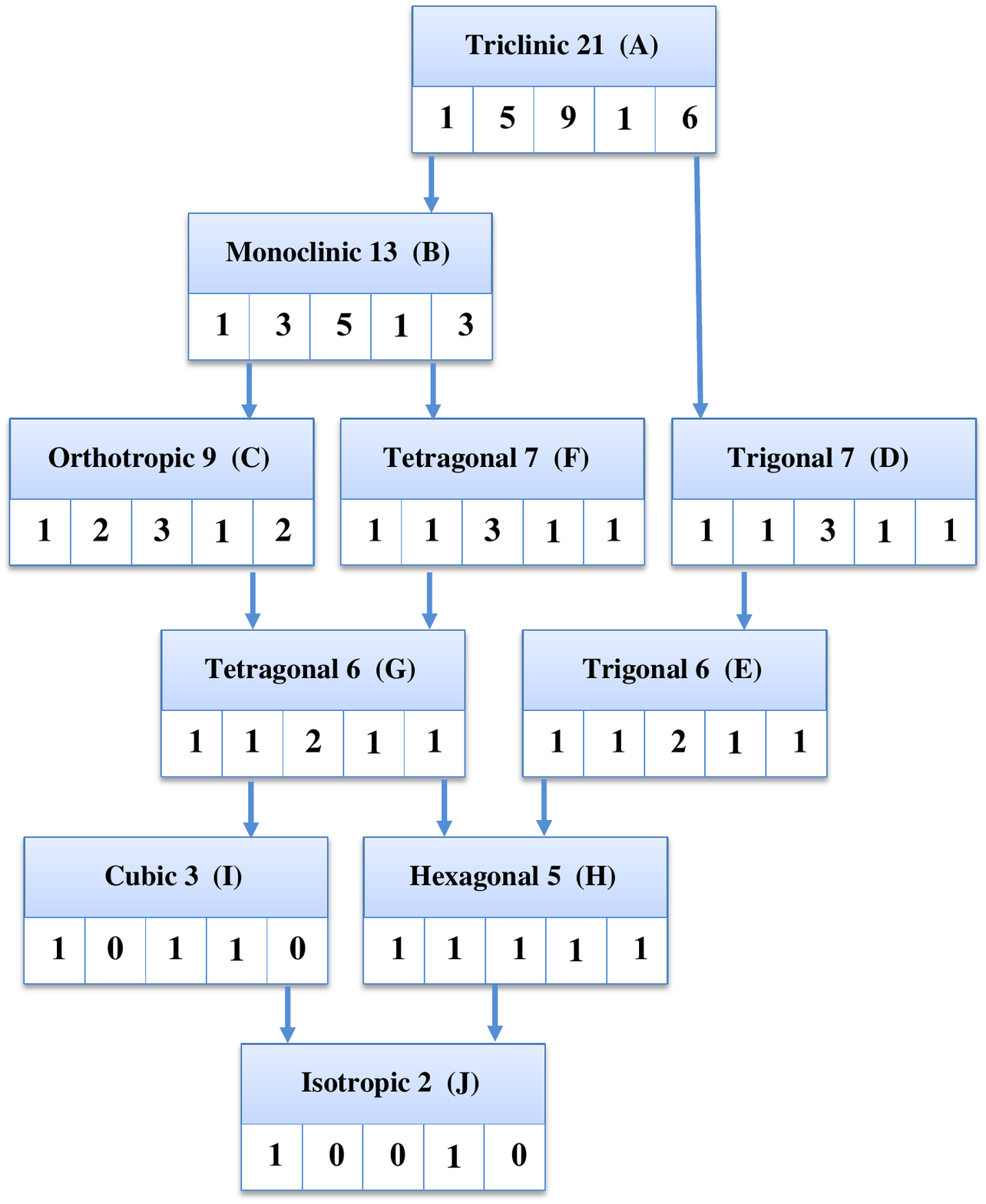,width=6.0in}}
\vspace*{-6.5cm}
\caption{Hierarchy of the symmetry systems. Dimensions of the tensor spaces and of their irreducible parts ${}^{(1)}S^{ijkl}$,   ${}^{(2)}S^{ijkl}$,  ${}^{(3)}S^{ijkl}$,  ${}^{(1)}A^{ijkl}$ and ${}^{(2)}A^{ijkl}$ are indicated.} 
\end{figure}

In order to have a more detailed description of the relation between the symmetry systems, we study now their inclusions and intersections. For briefness, we use now the notations $A,B,\cdots, J$ given in Tab. 1. For two systems connected in Fig. 1 by an arrow,    the lower system is included  into the higher one. Thus we have obvious inclusions 
\begin{equation}
    J\subset I\subset G\subset C\subset B\subset A \,,
\end{equation}
and so on. 
We consider now the relations between the system from different branches that are not directly connected one to another by an arrow. 
First we observe
\begin{equation}
    D\cap I=E\cap I=F\cap I=J\,.
\end{equation}
Moreover, 
\begin{equation}
    D\cap H=E\cap H=
    D\cap G=E\cap G=
    D\cap C=E\cap C=
    D\cap B=E\cap B= F
\end{equation}
and
\begin{equation}
    C\cap G=H
\end{equation}
Consequently, we have a diagram in Fig. 2 presented the inclusions of the symmetry systems in the form of Venn's diagrams from set theory. 

\begin{figure}[ph]
\vspace*{-8.5cm}
    \centerline{\psfig{file=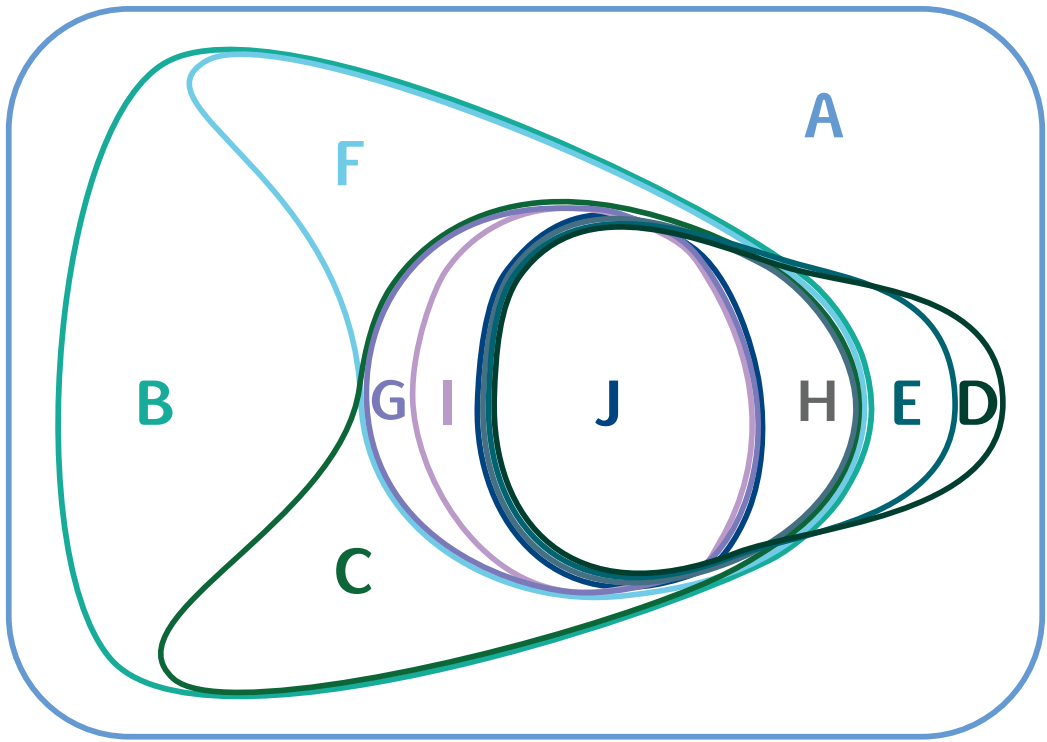,
    width=9.0in}}
\vspace*{-8.5cm}
\caption{Venn diagrams for inclusion of symmetry systems.} 
\end{figure}

\section{Conclusion}
In this paper,we presented matrix representations of the elasticity tensor that  confirm with its irreducible decompositions. In particular, $(3\times 7)$ matrix corresponds to $GL(3)$ decomposition of the elasicity tensor intoCauchy and non-Cauchy parts.  When the traces of the elasticity tensor are applied, an additional $SO(3)$ irreducible decomposition emerges.  We describe this decomposition with three $(3\times 3)$ matrices. Two of these  matrices are symmetric and one is of a  general form. Since the symmetric matrices represent two second  order tensors, their traces are extracted in an invariant form and generate two linear invariants of the  elasticity tensor. 

We apply the irreducible matrix decomposition  to all symmetry classes and decompose correspondingly their elasticity tensors. This resolution of the symmetry classes yields a natural scheme of the hierarchy and inclusion of the symmetry classes. This result is in an almost completely correspondence with the diagrams presented in literature. 

To our opinion the central problem related to the irreducible decomposition is the physical meaning of the independent pieces. We observe some preliminary results that can be derived from the resolution of the symmetry classes:
\begin{itemize}
    \item The scalar parts ${}^{(1)}S^{ijkl}$ and ${}^{(1)}A^{ijkl}$ with the scalars $S$ and $A$ enter all symmetry systems. Thus they can be considered as basic states of the deformed material. In particular, the closed isotropic prototype of a material can be derived by extracting these two scalars from the elasticity tensor. In algebraic description it means the orthogonal projection into two one dimensional subspaces, see \cite{Quad}. 
    \item The part ${}^{(3)}S^{ijkl}$ enclosed in all anisotropic crystal systems. Hence its norm can be used  as a  basic factor  of anisotropy. This term, however, enters only the Cauchy part. Thus without additional tensor parts, the non-Cauchy part is left isotropic. 
    \item Two tensor parts ${}^{(2)}S^{ijkl}$ with the tensor $P^{ij}$ and ${}^{(2)}A^{ijkl}$ with the tensor $Q^{ij}$ do not enter only the cubic and the isotropic systems. They can be used as additional characteristics of anisotropy. For trigonal, tetragonal and transverse isotropic systems, these parts are one-dimensional and can be completely described by their norms.

    \item For the systems with the symmetry higher than orthotropic, an invariant relation between second order tensors holds: the tensors $P^{ij}$ and $Q^{ij}$ are proportional to one another
    \begin{equation}
        P^{ij}=CQ^{ij}
    \end{equation} 
    with some numerical coefficient $C$. 
\end{itemize}

\end{document}